\documentclass[emulateapj,twocolumn]{aastex63}
\usepackage{graphicx}
\usepackage{float}
\usepackage{amsmath}
\usepackage{subfigure}
\usepackage{natbib}
\usepackage{array}
\usepackage{gensymb}
\usepackage{float}
\restylefloat{table}

\usepackage{booktabs}
\usepackage{xcolor}

\begin{document}


\title{Resolved UV and optical color gradients reveal environmental influence on galaxy evolution at redshift z$\sim$1.6}
\author{Cramer, W. J.}
\affil{Department of Physics and Astronomy, Notre Dame University, South Bend, IN 46617, USA}
\affil{School of Earth and Space Exploration, Arizona State University, Tempe, AZ 85287, USA}
\author{Noble, A. G.}
\affil{School of Earth and Space Exploration, Arizona State University, Tempe, AZ 85287, USA}
\affil{Beus Center for Cosmic Foundations, Arizona State University, Tempe, AZ 85287-1404, USA}
\author{Rudnick, G.}
\affil{Department of Physics \& Astronomy, University of Kansas, 1251 Wescoe Hall Drive, Malott room 1082, Lawrence, AD 66045}
\author{Pigarelli, A.}
\affil{School of Earth and Space Exploration, Arizona State University, Tempe, AZ 85287, USA}
\author{Wilson, G.}
\affil{Department of Physics and Astronomy, University of California, Riverside, 900 University Avenue, Riverside, CA 92521, USA}
\author{Bah{\'e}, Y. M.}
\affil{Laboratory of Astrophysics, Ecole Polytechnique F\'{e}d\'{e}rale de Lausanne (EPFL), Observatoire de Sauverny, 1290 Versoix, Switzerland}
\affil{Leiden Observatory, Leiden University, PO Box 9513, NL-2300 RA Leiden, the Netherlands}
\author{Cooper, M. C.}
\affil{Department of Physics \& Astronomy, University of California, Irvine, 4129 Reines Hall, Irvine, CA 92697, USA}
\author{Demarco, R.}
\affil{Institute of Astrophysics, Facultad de Ciencias Exactas, Universidad Andr{\'e}s Bello, Sede Concepci{\'o}n, Talcahuano, Chile}
\author{Matharu, J.}
\affiliation{Cosmic Dawn Center (DAWN), Denmark}
\affiliation{Niels Bohr Institute, University of Copenhagen, Jagtvej 128, DK-2200 Copenhagen N, Denmark}
\author{Miller, T. B.}
\affil{Center for Interdisciplinary Exploration and Research in Astrophysics (CIERA), Northwestern University, 1800 Sherman Avenue, Evanston, IL 60201, USA}
\author{Muzzin, A.}
\affil{Department of Physics and Astronomy, York University, 4700 Keele St., Toronto, Ontario, M3J 1P3, Canada}
\author{Nantais, J.}
\affil{Facultad de Ciencias Exactas, Departamento de Ciencias F{\'i}sicas, Instituto de Astrof{\'i}sica, Universidad Andr{\'e}s Bello, Fern{\'a}ndez Concha 700, Las Condes, RM 7591538, Chile}
\author{Sportsman, W.}
\affil{Department of Physics \& Astronomy, University of Kansas, 1251 Wescoe Hall Drive, Malott room 1082, Lawrence, AD 66045}
\author{van Kampen, E.}
\affil{European Southern Observatory, Karl-Schwarzschild-Str. 2, 85748, Garching bei M{\"u}nchen, Germany}
\author{Webb, T. M. A.}
\affil{Department of Physics, McGill Space Institute, McGill University, 3600 rue University, Montr{\'e}al, Qu{\'e}bec, Canada, H3A 2T8}
\author{Yee, H. K. C.}
\affil{The David A. Dunlap Department of Astronomy and Astrophysics, University of Toronto, 50 St George St., Toronto, ON M5S 3H4, Canada}

\begin{abstract}

The changes in colors across a galaxy are intimately connected to the galaxy's formation, growth, quenching history, and dust content. A particularly important epoch in the growth of galaxies is near $z \sim 2$ often referred to as `cosmic noon', where galaxies on average reach the peak of their star formation. We study a population of 125 cluster galaxies at $z \sim 1.6$ in three Hubble Space Telescope (\textit{HST}) filters, F475W, F625W, and F160W, roughly corresponding to the rest-frame FUV, NUV, and r band, respectively. By comparing to a control sample of 200 field galaxies at similar redshift, we reveal clear, statistically significant differences in the overall spatially resolved colors and color gradients in galaxies across these two different environments. On average, cluster galaxies have redder UV colors in both the inner and outer regions bounded by $r_{\mathrm{50}}$, as well as an overall wider dispersion of outside-in color gradients. The presence of these observed differences, along with evidence from ancillary data from previous studies, strongly suggests that the environment drives these population-level color differences, by affecting the stellar populations and/or dust content.

\end{abstract}

\section{Introduction}

A major question in astronomy is the role that the environment a galaxy lives in plays on its evolution. At low redshift, the interactions between galaxies in dense environments, as well as the interactions of galaxies with the gas between them, significantly change the population of observed galaxies. Overall, galaxies in clusters are more likely, when compared to groups and the field, to be red and quenched than blue and star forming (e.g. \citealp{Gunn+72, Dressler+80, Poggianti+99, Abadi+99, Boselli+06, Tonnesen+07, Wetzel+12, Brown+17, Jian+18, Davies+19}).

There are a variety of potential factors that drive these differences in cluster vs. field galaxy populations. These include, but are not limited to, starvation driven by interaction with the hot intercluster gas (e.g. \citealp{Kawata+08, McCarthy+08, Vijayaraghavan+15}), overconsumption of gas within galaxies due to high rates in cluster galaxies of galactic fountains \citep{Rasmussen+06, Bahe+15}, AGN \citep{George+19, Radovich+19}, and hydrodynamical interaction with the cluster environment driving gas inward \citep{Akerman+23}. Furthermore, a significant fraction of galaxies entering clusters experience pre-processing in the enhanced density environment (compared to the field) around clusters due to interaction with surrounding gas (e.g. \citealp{Tonnesen+07, Gabor+15, Lotz+19}) and/or enhanced gravitational interaction and merger rates due to the proximity to other galaxies (e.g. \citealp{Lotz+13, Hine+16, Coogan+18, Deger+18, Watson+19}). Additional detail and information on these effects and the state of our understanding of their impact in clusters is summarized in a review paper by \citet{Alberts+22a}.

One particularly commonly invoked driver of the population difference in clusters vs. the field at low redshift is ram pressure stripping \citep{Abadi+99, Boselli+06, Tonnesen+07}. As ram pressure strips gas (the material for replenishing star formation) from the outside-in, quenching driven by ram pressure would also be expected to follow this path. Indeed, several high-resolution studies of the stellar populations in individual galaxies at low redshift have found positive color gradients (redder outskirts than inner region) consistent with this radial trend in quenching \citep{Pappalardo+10, Abramson+11, Merluzzi+16, Fossati+18, Cramer+19}. However, during active stripping, observations of large samples of jellyfish galaxies in clusters have found overall star formation rates are enhanced when compared to the main sequence at low redshift (e.g. \citealp{Vulcani+18, Roberts+20, Boselli+21, Cramer+21, Durret+21, Roberts+22}). A similar enhancement was also recently found in jellyfish galaxies in clusters out to $z=0.55$ compared to field galaxies \citep{Vulcani+23}.

Field galaxy color gradients are much more extensively studied than those of cluster galaxies. At low redshift, both quiescent and star forming galaxies have been found to be much more likely to have bluer outskirts than the inner region (i.e. a negative color gradient) rather than redder outskirts than the inner region (a positive color gradient) \citep{Kormendy+89, Wu+05, Tortora+11}. Recent studies with \textit{HST} have also found similar negative color gradients for all types of field galaxies beyond $z>1$ \citep{Szomoru+13, Liu+16, Liu+17}. While the conclusions of these studies are limited by the difficulty of disentangling the degenerate effects of dust and quenching on optical and UV colors, galaxies at this redshift are primarily thought to grow from the inside-out \citep{Nelson+16, Hagen+16, Spilker+19}, supporting the hypothesis that negative color gradients may be connected to star formation. Certain combinations of colors can be used to disentangle the dust-quenching degeneracy directly, most popularly using the \textit{UVJ} diagram, which has been used to distinguish between star forming and quiescent galaxies at high-redshift \citep{Wuyts+07, Williams+09, Brammer+09, Whitaker+11, Muzzin+13b}. 

Arguably, the most impactful epoch in the evolutionary history of galaxies on a cosmic scale is $z \sim 2$, sometimes referred to as ``cosmic noon'', where galaxies reach the peak of their star formation \citep{Shapley+11, Madau+14}. However, to study color gradients and distinguish between outside-in and inside-out growth via star formation at high redshift requires observations that include the rest-frame near infrared. Until recently becoming available with JWST, these observations have been limited by resolution and sensitivity.  The only previously available instrument able to even access the rest-frame $J$-band at high-redshift was \textit{Spitzer}/IRAC, and that instrument had very poor resolution for use on high redshift galaxies. A recent study with JWST found that, of a sample of 54 star-forming field galaxies from $1.7 < z < 2.3$, 85\% of these galaxies had color gradients consistent with bluer outskirts than the inner galaxy and 70\% of the sample had this negative color gradient driven primarily by dust \citep{Miller+22}. This was the first data presented on resolved \textit{UVJ} color gradients of field galaxies at or near cosmic noon.

There are few to no large-scale published studies of the resolved color gradients of galaxies in clusters at any redshift. It is unclear at what epoch in the buildup of galaxy clusters ram pressure becomes a significant influence on galaxy evolution. Simulations have predicted that the ICM becomes built up enough for ram pressure to become significant for galaxy evolution in clusters around $z\sim2$ (see the recent review by \citealt{Boselli+22}). However, at high redshift, direct detection of `jellyfish' like ram pressure stripped tails seen at low redshift (e.g. \citealt{Yagi+10, Sun+10, Kenney+14, Cramer+19, Jachym+19}) is difficult. In contrast with other environmental quenching mechanisms discussed previously, there are a number of published analyses of the progression of an outside-in stellar population color gradient over time in ram pressure stripped galaxies \citep{Pappalardo+10, Abramson+11, Merluzzi+16, Fossati+18, Cramer+19}. The color effects of other quenching mechanisms are much less simple or constrained, and we lack clear observational studies of them equivalent to that done with ram pressure stripped galaxies. Thus, depending on the strength and stage of stripping, we could expect a study of cluster galaxies of sufficient sample size may find observable color differences compared to similar field galaxies, which are generally free of the influence of ram pressure.  Therefore, a population wide difference in color gradients between cluster and field could be a powerful tool for constraining the development and the role of the cluster environment in driving galaxy properties and evolution near cosmic noon. In order to conduct such an investigation, observations of high enough resolution to study gradients at this redshift are needed, especially with filters that can trace young star formation directly, such as in the UV.

As such, in this paper we present new UV observations of galaxies within three $z \sim 1.6$ clusters. The environmental quenching efficiency in these clusters, based on the number of non-star forming galaxies and defined as $f_{\text {eqe }}=\left(f_{\text {passive,cluster }}-f_{\text {passive, field }}\right) /\left(f_{\text {star-forming, field }}\right)$, is estimated to be $16 \pm 16\%$. Although there is significant uncertainty, this suggests that environmental quenching could be present, but may be low in these clusters at this redshift \citep{Nantais+17}. 

Later observations of these clusters have confirmed the presence of galaxies actively experiencing environmental influence. \citet{Noble+17, Noble+19} observed $\sim$25 galaxies across these three clusters with ALMA, and found asymmetric gas morphology and overall elevated gas fractions, suggestive of the influence of ram pressure. Indeed, studies that identified a number of galaxies affected by ram pressure also observed enhanced molecular gas fractions (M$_{\mathrm{H_2}}$ / M$_\star$) in the disk when compared to a control sample of field galaxies, often preferentially on the leading side of the galaxy, i.e. the side experiencing maximum ram pressure \citep{Jachym+14, Jachym+17, Verdugo+15, Lee+17, Moretti+18, Cramer+20, Cramer+21, Roberts+22, Vulcani+23}. 

In \citet{Cramer+23}, we conducted a follow-up kinematic study of this sample of ALMA detected $z \sim 1.6$ cluster galaxies and found that the overall rates and degrees of kinematic asymmetry in the molecular gas component were dramatically elevated when compared with field galaxies at similar redshift \citep{Cramer+23}. While these asymmetries cannot be incontrovertibly attributed to ram pressure stripping because of the limited sensitivity and resolution resulting in only a few tenuous detections of stripped tails, the observed kinematic asymmetry patterns in the disk are extremely similar to those observed in low-redshift jellyfish galaxies \citep{Bacchini+23}. Combining the evidence of a moderate elevation in quenching with the direct evidence of ram pressure stripping, suggests that these clusters supports that these clusters are in early stages of environmentally driven quenching. They are thus ideal candidates for a follow-up \textit{HST} study of resolved color gradients which is the subject of this work.

\section{Observations}

The three clusters of galaxies studied in this paper, J022426–032330 (J0224), J033057–284300 (J0330), and J022546–035517 (J0225), were discovered within the 42 deg$^2$ Spitzer Adaptation of the Red-sequence Cluster Survey (SpARCS) fields (see additional information in Table 1 of \citealt{Nantais+16}). All three clusters are spectroscopically confirmed with redshifts of $z=1.63$, $z=1.63$, and $z=1.59$, respectively \citep{Lidman+12, Muzzin+13, Nantais+16}, and contain, in total, 91 spectroscopically confirmed members within $\delta_{\mathrm{zspec}} = 0.015$ of the cluster BCG \citep{Nantais+16, Nantais+17}. In total, there are observations over 16 bands spanning optical to NIR ($ugrizYK_{s}$ and F160W) as well as IR and FIR (3.6/4.5/5.8/8.0/24/250/350/500$\mu$m), that are used to estimate photometric redshifts with \texttt{EAZY} for the entire cluster field \citep{Wilson+09, Muzzin+09, Muzzin+13, Nantais+16}. The photometric cluster membership criterion is $(z_{\text{phot}} - z_{\text{cluster}})/(1 + z_{\text{cluster}}) \leq 0.05$, which for these clusters is approximately $1.5 < z_{\text{phot}} < 1.76$. 

\subsection{\textit{HST} Observations}

The central cluster regions have deep \textit{HST} imaging with the F160W filter from the WFC3-IR camera from the “See Change” program (GO-13677 and GO-14327; \citealt{Hayden+21}). Furthermore, the clusters were also imaged with the F475W (rest-frame FUV) and F625W (rest-frame NUV) filters with ACS (GO-16300, PI: Noble). Observations in each of these two filters consisted of data from 9 total executions across 3 orbits, for a total of 6882 seconds of exposure time in J0225 and J0330. Due to lost guiding during individual exposures in some orbits of the observations of J0224, we applied for and received additional orbits to make up for the lost exposure time. This resulted in the F475W and F625W images of J0224 having a total exposure time of 8319s and 6579s respectively. The data were reduced using the \texttt{Drizzlepac} software by aligning the individual frames using \texttt{tweakreg}, and drizzling the aligned frames using \texttt{astrodrizzle} with a pixel size of 0.035{\arcsec} for F475W and F625W, and 0.05{\arcsec} for F160W. The images were then matched to a common grid using the Astromatic software \texttt{SWarp} \citep{Bertin+02}, to allow for pixel-by-pixel analysis and comparison.\footnote{All the \textit{ HST} data from these clusters used for this paper can be found in MAST: \dataset[10.17909/40tq-8k79]{http://dx.doi.org/10.17909/40tq-8k79}.} In Figure \ref{fig:optical_image} we show a color composite image with the reduced observations from \textit{HST} F475W, F625W, and F160W filters of the J0224 cluster.

The images were then PSF-matched based on empirically created effective point spread functions (ePSFs) based on the selected point-like stars in the image with the \texttt{photutils} \citep{photutils} Python package. Stars in each cluster image were visually inspected in each filter to ensure a clean final ePSF. The J0224 ePSF was created from 9 stars, J0225 from 8 stars, and the 3 stars in the J0330 image were not sufficient to create an ePSF. We chose to use the J0224 ePSF for the J0330 image matching since it was derived from the most stars. The impact of this choice should be minimal as the J0224 PSF stars in the images failed to strongly capture the diffraction spikes of the secondary mirror, so the derived PSF is close to a Gaussian. Furthermore, the position angle of the J0224 and J0330 observations is similar, so any diffraction spikes would have been in similar locations. We employed the \texttt{pypher} \citep{Boucaud+16} Python package to create matching kernels, to redistribute the flux from one band of imaging to match the flux distribution of the target band of imaging, with a regularization parameter of 10$^{-4}$. The images were then convolved using the \texttt{convolve} method from \texttt{astropy}.

\subsection{Hubble Frontier Fields Catalogues}

In order to have as close a comparison sample of field galaxies as possible at a similar redshift as the $z\sim1.6$ clusters, we required a dataset with the same, or at least similar filters. Furthermore, we also needed high-resolution observations for a resolved study of radial color gradients, and relatively deep observations to allow for a large comparison sample. The HFF-DeepSpace (HFF-DS) program, data release v3.9 \citep{Shipley+18}, provides the best sample for satisfying these three criteria. The program includes 6 parallel pointings of the outskirts of galaxy clusters near $z \sim 0.5$. The high redshift galaxies that fall in these fields represent an unbiased field galaxy sample, with no significant redshift clustering at these redshifts indicative of any chance contamination with high redshift cluster galaxies \citep{Shipley+18}. The imaging with ACS includes observations with the F435W, F606W, and F160W filters. In total, the program includes 840 orbits of Director's Discretionary time, resulting in extremely deep imaging which allows us to build a large sample of 200 $z \sim 1-2$ field galaxies in the parallel fields. \footnote{The reduced, PSF-matched images that we use are publicly available upon request at \url{http://cosmos.phy.tufts.edu/~danilo/HFF/Download.html}.}

\section{Methodology}

Our goal is to study the radial colors and color gradients of high-z cluster galaxies and compare them to the field, in order to search for population level differences in the two samples. In the following sections, all magnitudes are given in units of AB magnitudes.

\subsection{Image Processing and Source Identification}

While the \textit{HST} reduction process includes sky subtraction, some residual sky was still evident in the F475W and F625W images from the high-$z$ cluster data. We applied a Gaussian smoothing kernel with a size of 3 pixels, and a threshold of 3$\sigma$ to a source subtracted image (based on a segmentation map from Source Extractor \citep{Bertin+96}), and subtracted this smoothed image from the image data, in order to account for this background. The field data available from the HFF-DS sample had already been background subtracted in the same manner. We then identify all sources above a threshold of 5$\sigma$, and deblend them using \texttt{photutils.deblend\_sources} \citep{photutils}. Deblending was done with \texttt{nlevels=32} and \texttt{contrast=0.001}. We tested a range of these deblending parameters and visually inspected the resulting samples to hone in on the best values to use. We calculate the Kron flux and the major and minor axis of the ellipse encompassing the Kron flux.

We then cross match the cluster sample we detect within the F160W field of view with spectroscopic and photometric catalogues from \citet{Nantais+16} to isolate galaxies within the redshift range of the high-$z$ clusters ($1.5 < z < 1.76$). We find a total of 65 spectroscopically confirmed galaxies, and an additional 60 galaxies with photometric redshifts within this range. \citet{Nantais+16} found that the scatter of $(z_{\mathrm{phot}} - z_{\mathrm{spec}}) / (1+z_{\mathrm{spec}})$ excluding outliers is $\sigma=0.04$ for galaxies with a $z_{\mathrm{phot}}$ within $1.5 < z < 1.76$. The outlier rate within the cluster region, defined as the percentage of objects with $(z_{\mathrm{phot}} - z_{\mathrm{spec}}) / (1+z_{\mathrm{spec}}) \ge 0.15$, as defined in \citet{Burg+13} is 5.4\%. While spectroscopic redshifts are in general more reliably constrained than photometric redshifts, we find no statistically significant effect on any of our analysis from comparing the full sample we use with one with only spectroscopically confirmed cluster members (see Appendix and Figures \ref{fig:app_specz_colors} \& \ref{fig:app_specz_quadrant} for details). For the field sample, we select all galaxies between $1.2 < z_{\text{phot}} < 2.0$, a total of 200 galaxies; having a large sample of galaxies with which to compare is important for this analysis.

\subsection{Aperture flux extraction}

\begin{figure*}
	\plotone{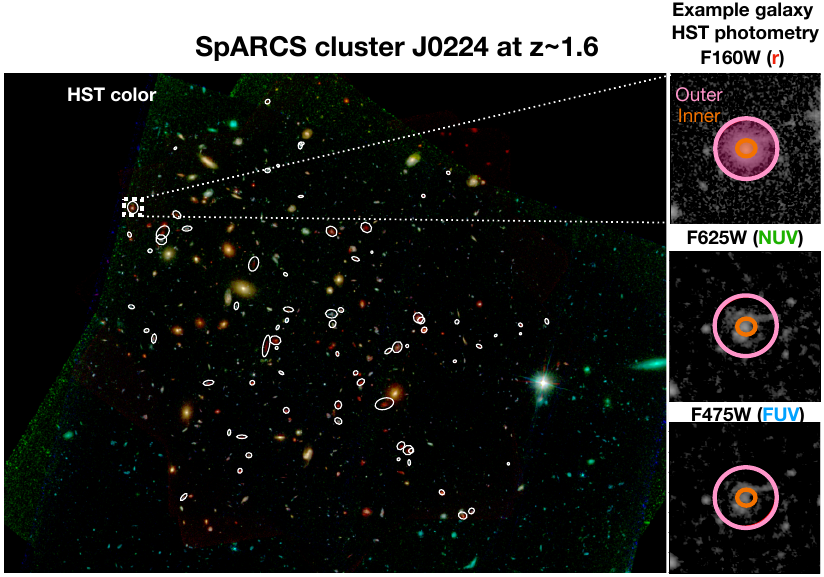}
	\caption{A composite color image of one of the three SpARCS clusters we study, J0224, with \textit{HST} imaging in filters F475W (blue), F625W (green), and F160W (red). Members of this cluster considered for this paper are indicated with white ellipse. On the right is an example of the inner aperture ($r < r_{50}$, the orange shaded region) and outer annulus ($r_{50} < r < r_{90}$, the pink shaded region) determined for this galaxy, used for the analysis presented in this paper. The size and ellipticity of the two apertures is determined based on the total flux in the F160W filter, and then applied to both the F625W and F475W images.}
	\label{fig:optical_image}
\end{figure*}

In order to study color gradients, we calculate the flux within an outer annulus and an inner aperture for each galaxy. To constrain radial evolution within a galaxy, it is especially important to effectively compare the flux within annuli with geometry consistent with the overall ellipticity of each galaxy. Commonly used source finding software packages like Source Extractor \citep{Bertin+96} and photutils can both be configured to output the parameters of a circular aperture containing a user specified fraction of the total flux, but not an elliptical aperture. We modified the photutils code to use an elliptical optimizing function, instead of a circular optimizing function, to fit proper elliptical apertures to each galaxy. We find the ellipses encompassing 50\%, and between 50\% and 90\%, of the total Kron flux in the F160W filter for each galaxy, then apply these same outer and inner ellipses to each accompanying filter pair. Thus, the inner and outer region of a galaxy in all three filters is measured with the same apertures, as set by the F160W surface brightness profile of each galaxy. An example of these apertures overlayed on a cluster galaxy from our sample is shown in Figure \ref{fig:optical_image}.

\subsection{Flux error estimation}

Determining the uncertainties for our measured fluxes is complicated by the correlated nature of the noise in our images that results from the sub-pixel drizzling process.  The measured pixel-to-pixel variation therefore underestimates the true uncertainty.  To address this, we make an empirical estimate of the uncertainties using the technique from \citet{Labbe+03} and \citet{Shipley+18}, which worAD under the assumption that the uncertainty in the measured fluxes is dominated by the background uncertainty, which is true for our images.  We refer the reader to those papers for a detailed description of the technique.  In summary, we placed empty apertures throughout each image, ensuring that the apertures do not overlap with detected objects.  For a given sized aperture, the distribution of the measured flux in these empty apertures has a Gaussian distribution.  The standard deviation of this distribution gives the flux uncertainty for apertures of identical area ($\sigma_f$).  This procedure was repeated for many different aperture sizes and a 2nd order polynomial was fit to the trend of $\sigma_f$ vs. aperture size.  As shown in \citet{Labbe+03}, this is a good fit to the trend, which we verified with our images.  For every aperture in which we measured a flux in our image we use these fits to find the uncertainty in apertures of identical area in pixels.  This becomes our quoted uncertainty.

\subsection{Sample matching}

The HFF-DS catalogue is the best sample for a field comparison in terms of depth of observations and thus the abundance of galaxies in our desired redshift range at \textit{HST} resolution. However, there are several key differences between our sample and that from HFF-DS, that need to be addressed for a proper comparison. The field sample of the HFF-DS survey and our data have slightly different filters (F435W \& F606W versus F475W \& F625W). Furthermore, to maximize the available sample of control galaxies, we include HFF-DS galaxies from $1.2 < z < 2.0$ while cluster galaxies are between $1.53 < z < 1.69$. In order to make the closest possible comparison between the two samples, we correct measured HFF-DS galaxy fluxes with both a filter and redshift correction. 

\begin{table*}[]
\begin{tabular}{@{}ccccc@{}}
\toprule
\textbf{ } & \textbf{Detected in} & \textbf{Non-detected} & \textbf{Non-detected} & \textbf{Non-detected} \\
 & \textbf{all three filters} & \textbf{only F475W} & \textbf{only F625W} & \textbf{F475W \& F625W} \\
\midrule
\textbf{\begin{tabular}[c]{@{}c@{}}Cluster galaxy\\ (inner region)\end{tabular}} & 94  & 22 & 1 & 8  \\
\textbf{\begin{tabular}[c]{@{}c@{}}Cluster galaxy\\ (outer region)\end{tabular}} & 90  & 20 & 4 & 11 \\
\textbf{\begin{tabular}[c]{@{}c@{}}Field galaxy\\ (inner region)\end{tabular}}   & 188 & 11 & 0 & 1  \\
\textbf{\begin{tabular}[c]{@{}c@{}}Cluster galaxy\\ (outer region)\end{tabular}} & 184 & 13 & 0 & 3 \\
\bottomrule
\end{tabular}
\caption{Number of detections in the inner and outer galaxy regions above $3\sigma$ in both the cluster and fields samples. The colors of these regions are plotted in Figure \ref{fig:color_diagram}.}
\label{tab:detections}
\end{table*}

The filter correction was done by analyzing  a series of both SSP and $\tau = 0.6$ Gyr models from \texttt{EzGal} \citep{Mancone+12} at a range of ages (see Figure \ref{fig:filter_conversion}). We estimate the predicted flux within the F435W, F475W, F606W, and F625W filters using \texttt{stsynphot} \citep{stsynphot}. We then fit the scatter plot of points consisting of the 22 different models with a quadratic function, which allows us to determine the conversion between filters (see Figure \ref{fig:filter_conversion}). We find the following relations:

\begin{align*}
(F435W - F475W) = 0.052 (F435W - F606W)^2 \\ + 0.19 (F435W - F606W) + 0.010
\end{align*}

\begin{align*}
(F606W - F625W) = -0.043 (F435W - F606W)^2 \\ + 0.21 (F435W - F606W) - 0.001
\end{align*}

The 1$\sigma$ errors on these fits are added in quadrature with the flux errors for each galaxy from the empty aperture method for the rest of this analysis. The filter correction is then adjusted for the redshift of the field galaxies, ranging from $z=1.2-2.0$ rounded to the nearest 0.1, once more using \texttt{stsynphot} to shift the template spectra to each redshift step before calculating the estimated filter flux conversions. 

After equalizing these filters, for simplicity we will hereafter refer to the these filters by their nearest overlapping rest-frame color band, the F475W filter as FUV, the F625W as NUV, and the F160W as \textit{r}. We note that the FUV and NUV bands have both been found to trace star formation within the last $\sim$100 Myr, with the FUV more sensitive than the NUV to star formation at even earlier times, within $\sim$10-40 Myr \citep{Abramson+11, Ephremova+11, Boqiuen+14}.

\begin{figure*}
	\plotone{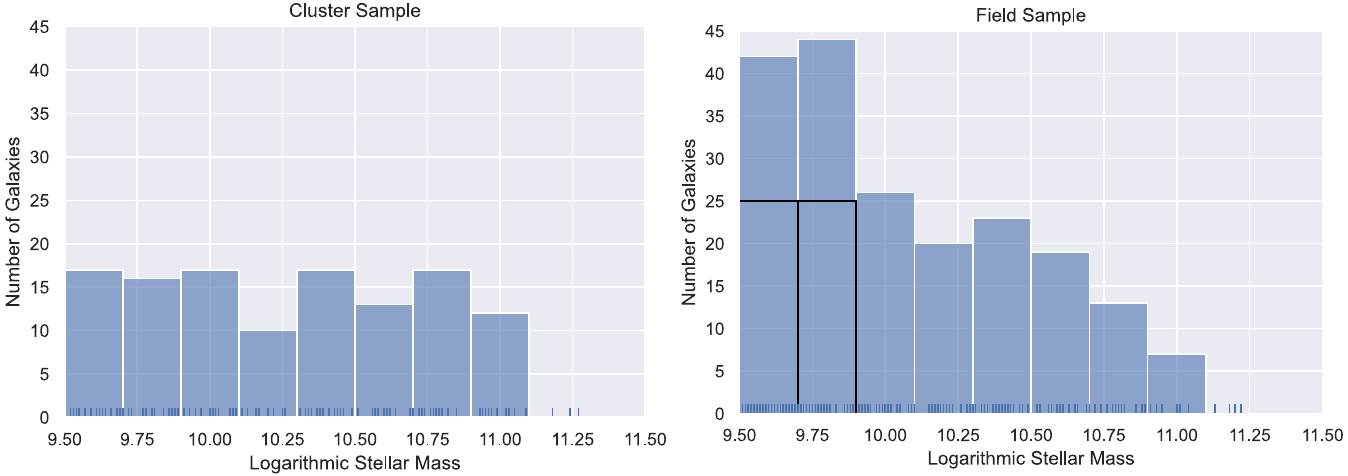}
	\caption{Stellar mass distributions for the SpARCS cluster sample based, on values from \citet{Nantais+16}, and the field sample, based on values from \citet{Shipley+18}. While the field sample has more lower mass galaxies, repeating the analysis with more closely matched stellar mass samples by reducing the number of galaxies in the firs two bins (shown in black) results in no statistically significant difference to the conclusions of this work.}
	\label{fig:filter_conversion}
\end{figure*}

We also impose a restriction on galaxy stellar mass in each sample to be between $M_{\star} = 10^9 - 10^{11}~{\rm M}_{\odot}$, based on estimates from \texttt{FAST} modelling \citep{FAST} presented in \citet{Nantais+16} for the cluster, and \citet{Shipley+18} for the field. As the HFF-DS observations are deeper than the cluster observations, which are mass limited to $M_{\star} \sim 10^9~{\rm M}_{\odot}$, this cut is necessary to make sure we are comparing similar types of galaxies. The stellar mass distributions of our samples are similar, but there are about two times as many galaxies of $\sim 10^9~{\rm M}_{\odot}$ in the field sample as the cluster, possibly due to the difference in the depth of the observations, or the larger total area covered by the HFF-DS survey. We re-ran our analysis, randomly removing low mass field galaxies until the total number in the low mass bins matched the number of galaxies in the cluster sample, and found no statistically significant difference in any of our results and conclusions.

The full details of the number of galaxies with fluxes detected in one, two, or all three filters are shown in Table \ref{tab:detections}.

\begin{figure*}
	\plotone{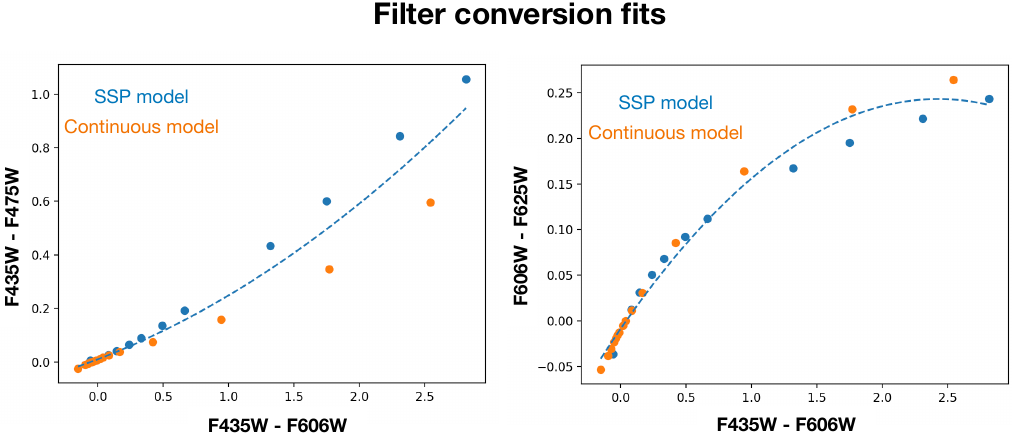}
	\caption{Quadratic fits to the predicted color from SSP and $\tau = 0.6$ Gyr models produced at ages of 0.01, 0.05, 0.1, 0.2, 0.3, 0.4, 0.5, 0.8, 1.0, 1.5, 2.0, 3.0, 4.0, 5.0, and 6.0 Gyr by \texttt{EzGal} using a BC03 IMF \citep{Chabrier+03} (see text for additional details). Axes are in units of AB magnitudes. With these fits, we establish a formula for converting the F435W to the F475W filter, and the F606W to the F625W filter, based on the observed F435W - F606W color, in order to directly compare fluxes between the cluster and field.}
	\label{fig:filter_conversion}
\end{figure*}

\section{Results}

\begin{figure*}
	\plotone{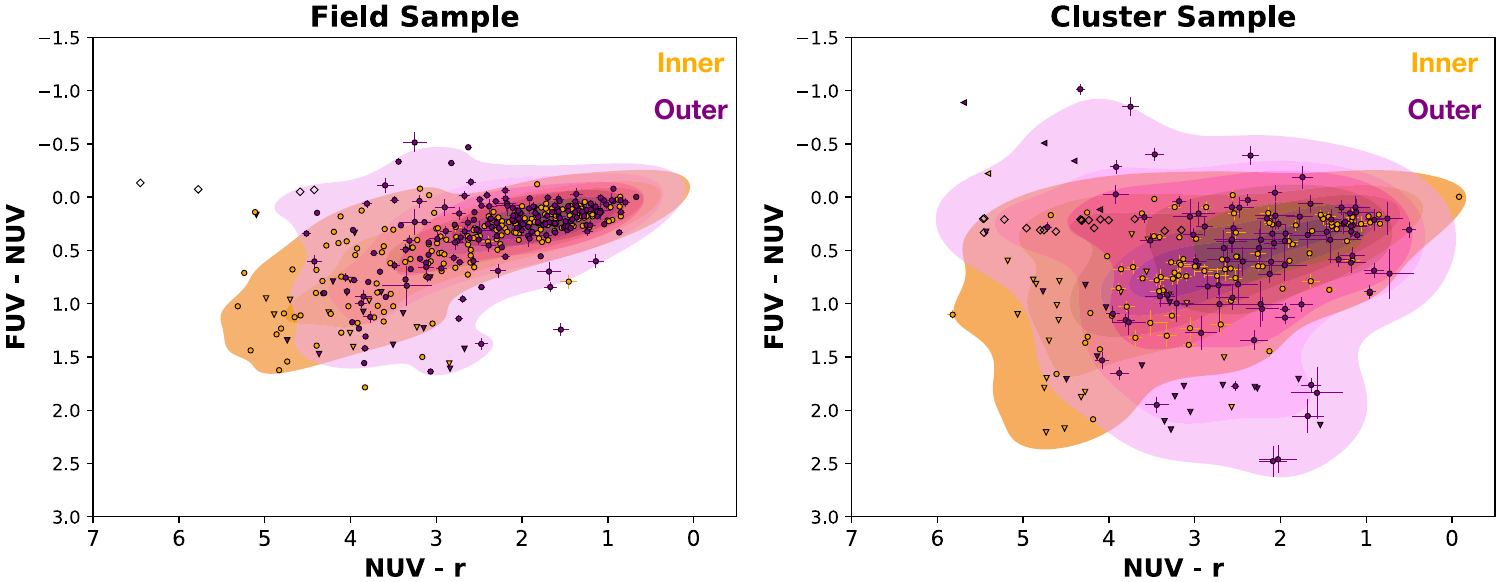}
	\caption{A comparison of the inner (orange) and outer (pink) NUV-r vs. FUV-NUV colors for the field sample and the cluster sample. Left facing triangles show aperture fluxes at $3\sigma$ limits in either the FUV flux, or the NUV flux. Downward facing arrows show limits in just the NUV flux. Open circles show aperture fluxes with limit fluxes in both FUV and NUV. Contours based on a kernel density estimate (KDE) are also shown to better illustrate the comparison. Six contour levels are shown, corresponding to $\sim 16\%$ of the total plotted point distribution per contour level. Overall, the field sample shows a tighter overall distribution, while the cluster sample shows a wider range of values in the FUV-NUV color. Overall error levels on colors are quite low, averaging $\sim$0.05 mags for the field sample, and $\sim$0.1 mags for the cluster sample.}
	\label{fig:color_diagram}
\end{figure*}

The result of the color analysis of both cluster and field galaxies is shown in Figure \ref{fig:color_diagram}. Simply by eye, it is apparent that field galaxies are overall more tightly distributed than cluster galaxies, with both inner and outer region colors centered between approximately 0 and 0.5 (FUV-NUV), and between 0.5 and 2.5 (NUV-r). The distribution of cluster galaxy colors, both inner and outer, is much more dispersed than field galaxies, although the largest concentration of points is also centered near FUV-NUV of 0.5 to 0, and NUV-r between 2.5 to 0.5. We calculate the variance ($
\sigma ^2$) of the cluster and field sample along each axis. The cluster sample, in the inner regions, has a variance of 0.5 in the FUV-NUV color, and 1.4 in the NUV-r color. In the contrast the field sample has a variance of  0.15 in the FUV-NUV color, and 1.0 in the NUV-r color. The cluster sample, in the outer regions, has a variance of 0.5 in the FUV-NUV color, and 1.0 in the NUV-r color. In the contrast the field sample has a variance of  0.15 in the FUV-NUV color, and 1.4 in the NUV-r color. These calculations show that the cluster sample is more dispersed along both axes than the field sample, and especially so when comparing the FUV-NUV colors in the outer galaxy regions. Both cluster and field galaxies have a similar total range of NUV-r values, ranging from approximately 6 to 0.

While these insights from visual inspection can be useful for assessing the potential differences between the distributions within the cluster and field samples, proper statistical analysis of these distribution is necessary in order to draw strong conclusions.

\subsection{Statistical Survival Analysis}

It is important to compare the overall properties of galaxies in the clusters we study with the matched field sample in a statistically sound manner. In order to do this, and understand the impact of limit cases in our statistical analysis, we use a method called survival analysis. Survival analysis originates from public health, and is commonly used to measure mortality over the course of a drug trial, where ``censored" data need to be considered to account for patients that leave the trial early. It can be used in our case to consider the color distributions of the cluster and field sources, while incorporating the galaxies and galaxy regions with only upper limits on the flux as censored data. While more well-known in the public health field, this analysis technique has been used for similar purposes to that we described here in a number of other astronomy papers \citep[e.g.][]{Mok+16, Odekon+16}. All galaxies in our samples are detected in r, but $\sim 20\%$ of the sample have an upper limit flux in FUV or NUV, that are important to consider, as limits still provide useful color constraints.

\begin{figure*}
  \begin{minipage}[b]{1.0\linewidth}
    \centering
	\plotone{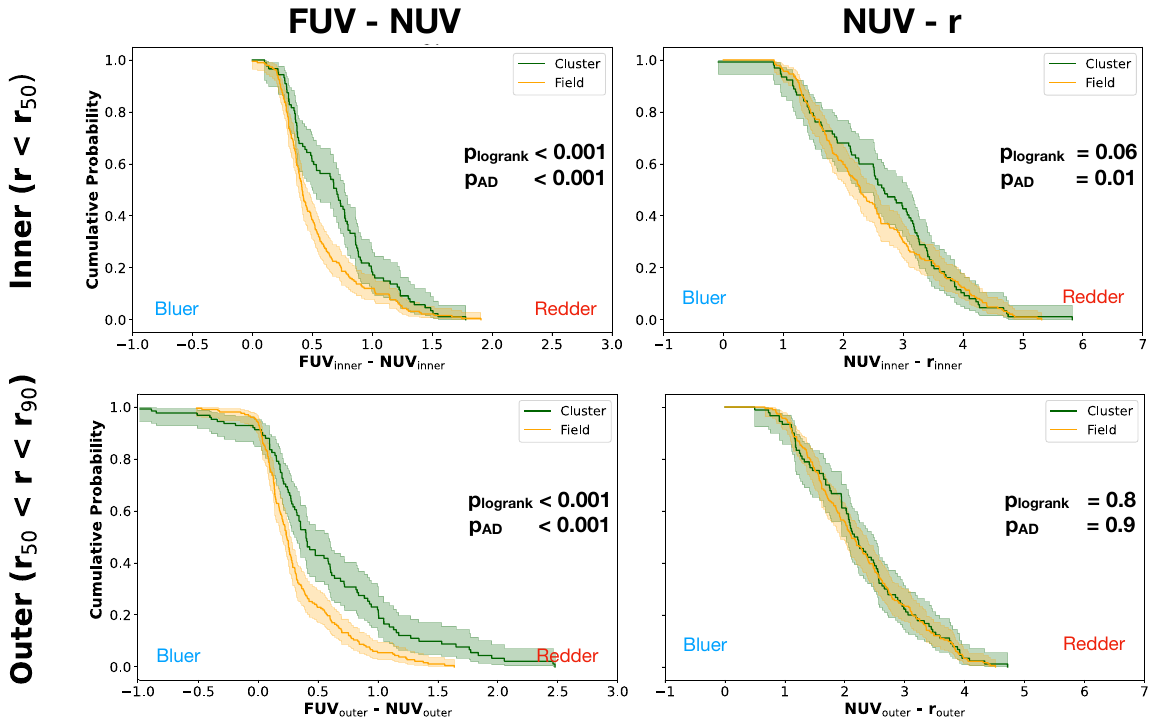}
	\caption{A comparison of the cumulative distribution function best fits to the cluster and field samples. The shaded regions show the 95\% confidence interval, which is calculated based on the sample size, sampling of the data, and the number of limit cases. The x-axis is in units of AB magnitudes. We show a comparison of the cumulative distributions of the inner regions (top) and the outer regions (bottom) for cluster (in green) and field (in orange) galaxies. For the FUV-NUV color, both the inner and outer comparisons show statistically significant differences based on the \textit{p}-value of a log-rank comparison. For the NUV-r color, the inner and outer comparisons show no statistically significant differences based on the \textit{p}-value.}
	\label{fig:FUV_survival}
  \end{minipage}%
  \begin{minipage}[b]{1.0\linewidth}
    \centering
	\plotone{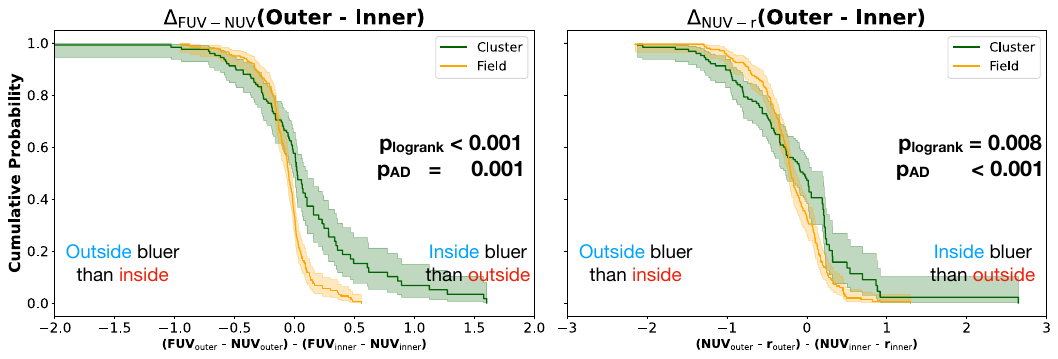}
	\caption{\textbf{Top:} A comparison of the cumulative distribution function fits to the cluster and field samples. The shaded regions show the 95\% confidence interval, which is calculated based on the sample size, sampling of the data, and the number of limit cases in the sample. The x axis is in units of AB magnitudes. We show a comparison of the cumulative distributions of the outer-inner color gradient distributions for cluster and field galaxies. For the FUV-NUV color (left), only the logrank test supports a different cluster vs. field gradient distribution, but nonetheless, a number of the cluster galaxies do have steeper redder outside to bluer inside trends. For the NUV-r color (right), the overall outer-inner distribution is statistically significantly different, but without a clear bias towards more negative or positive gradients.}
	\label{fig:NUV_survival}
  \end{minipage}
\end{figure*}

We fit a survival statistics Kaplan-Meier function, a non-parametric, maximum likelihood statistical estimator, to both the field and cluster samples for each color. We then use a log-rank test to assess the similarity in the distribution of both functions, and calculate a \textit{p}-value to estimate whether the two populations are significantly different. The survival statistics method and the logrank test do not incorporate measurement errors directly. We are able to calculate the 95\% confidence interval (shown in Figure \ref{fig:FUV_survival}), based on the sample size, sampling of the data, and the number of limit cases, but not the measurement errors. In order to assess the impact of these errors on the \textit{p}-value we calculate, we implement a Monte-Carlo approach, using the 1$\sigma$ errors from our empty aperture analysis. We vary each filter component of each color within a Gaussian distribution, with a standard deviation equal to the 1$\sigma$ error of each flux value, then proceed with a Kaplan-Meier fit, and calculate a \textit{p}-value with a log-rank method, over 100 iterations. This allows us to estimate the variance of the \textit{p}-value within the 1$\sigma$ flux errors. We find that the individual flux errors do not contribute to a significant change in the overall confidence level of the \textit{p}-values we calculate. Plots showing the Monte-Carlo generated Kaplan-Meier fits can be found in the Appendix (Figures \ref{fig:app_survival_f1} \& \ref{fig:app_survival_f2}).

\begin{table*}[]
\begin{tabular}{@{}ccccccc@{}}
\toprule
\textbf{ } & \textbf{Cluster Median} & \textbf{Field Median} & \textbf{Logrank} & \textbf{Logrank} & \textbf{AD Test} \\
 & \textbf{Color} & \textbf{Color} & \textbf{P-value} & \textbf{P-value STD} & \textbf{P-value}\\
\midrule
FUV - NUV (inner) & 0.6 & 0.3 & $<$0.001 & $<$0.001 & $<$0.001 \\
FUV - NUV (outer) & 0.4 & 0.2 & $<$0.001 & $<$0.001 & $<$0.001 \\
$\Delta$ FUV - NUV & 0.02 & -0.04 & $<$0.001 & $<$0.001 & 0.001 \\
NUV - r (inner) & 3.1 & 2.4 & 0.06 & 0.02 & 0.01 \\
NUV - r (outer) & 2.1 & 2.1 & 0.8 & 0.1 & 0.9 \\
$\Delta$ NUV - r & -0.04 & -0.2 & 0.008 & 0.02 & $<$0.001 \\
\bottomrule
\end{tabular}
\caption{Results of survival analysis of field and cluster galaxy samples' colors. From left to right, in column (1) the median value of the color for cluster galaxies, in column (2) the median value of the color for field galaxies. In column (3) the \textit{p}-value from a logrank comparison, and in column (4), the standard deviation in the \textit{p}-value based on a Monte-Carlo approach for estimating the effect of the flux errors on the CMDs (described further in the Appendix). In column (5) are the \textit{p}-values resulting from an Anderson-Darling test between the curves, considering only the uncensored data points.}
\label{tab:p-values}
\end{table*}

We compare the cumulative distribution functions (CDFs), derived from the survival statistics methods, of the cluster and field galaxies in two galaxy regions, from $r < r_{50}$, and from $r_{50}$ to $r_{90}$. These are the inner and outer FUV-NUV color, the inner and outer NUV-r color, and the color radial gradient from subtracting FUV$_{\mathrm{outer}}$-NUV$_{\mathrm{outer}}$ from FUV$_{\mathrm{inner}}$-NUV$_{\mathrm{inner}}$, as well as NUV$_{\mathrm{outer}}$-r$_{\mathrm{outer}}$ from NUV$_{\mathrm{inner}}$-r$_{\mathrm{inner}}$. The results are shown in Figures \ref{fig:FUV_survival} \& \ref{fig:NUV_survival}.

It has been found that the results of the logrank test can be less certain when comparing two survival curves that cross each other \citep{Lin+09, Dormuth+22}. While an analysis of a large number of published papers that have data with crossing survival curves found a significant fraction still use the logrank test in this situation \citep{Li+15}, we wanted to be careful that our results were not significantly impacted by this possibility. For this reason, we also used the Anderson-Darling (AD) test to compare the cluster and field color distributions (the limitation of this approach is that it does not incorporate the flux limit cases). The results of the logrank test and the AD tests are shown in Table \ref{tab:p-values}.

For the logrank and AD test of the FUV-NUV color distributions we find \textit{p}-values of less than 0.001 for both inner and outer regions, indicating a $>3\sigma$ level difference between the distributions of cluster and field galaxy UV colors. Furthermore, cluster galaxies are redder than field galaxies in both outer and inner regions, based on the median UV color of each population. In contrast, in both outer and inner regions, there is no significant difference in the NUV-r color between field and cluster galaxies as seen in Figure \ref{fig:FUV_survival} (right). 

The CDF comparison of the color gradient in  (NUV$_{\mathrm{outer}}$-r$_{\mathrm{outer}}$) - (NUV$_{\mathrm{inner}}$-r$_{\mathrm{inner}}$) indicates a statistically significant difference in color gradient distribution when comparing cluster to field galaxies. The color gradient for (FUV$_{\mathrm{outer}}$-NUV$_{\mathrm{outer}}$) - (FUV$_{\mathrm{inner}}$-NUV$_{\mathrm{inner}}$) appears to show statistically significant differences between the two samples according to both the logrank test, and the AD test. Along with different overall distributions, the range of the FUV-NUV gradients in clusters indicates the cluster sample contains galaxies with steeper outside-in reddening gradients than the field.

Understood together, these results indicate that cluster galaxies are more likely to have redder inner and outer regions in the UV than field galaxies, and redder inside-out color gradients across cluster galaxies. In contrast, cluster and field galaxies have statistically similar overall NUV-r colors in the inner and outer regions. The results on the NUV-r color gradients suggest a difference between the two populations, although the p-value for the survival statistics test is significantly larger than that for the FUV-NUV color gradients (0.008 vs $<0.001$ respectively).

\subsection{Quadrant Analysis}

\begin{figure*}
  \begin{minipage}[b]{1.0\linewidth}
    \centering
    \plotone{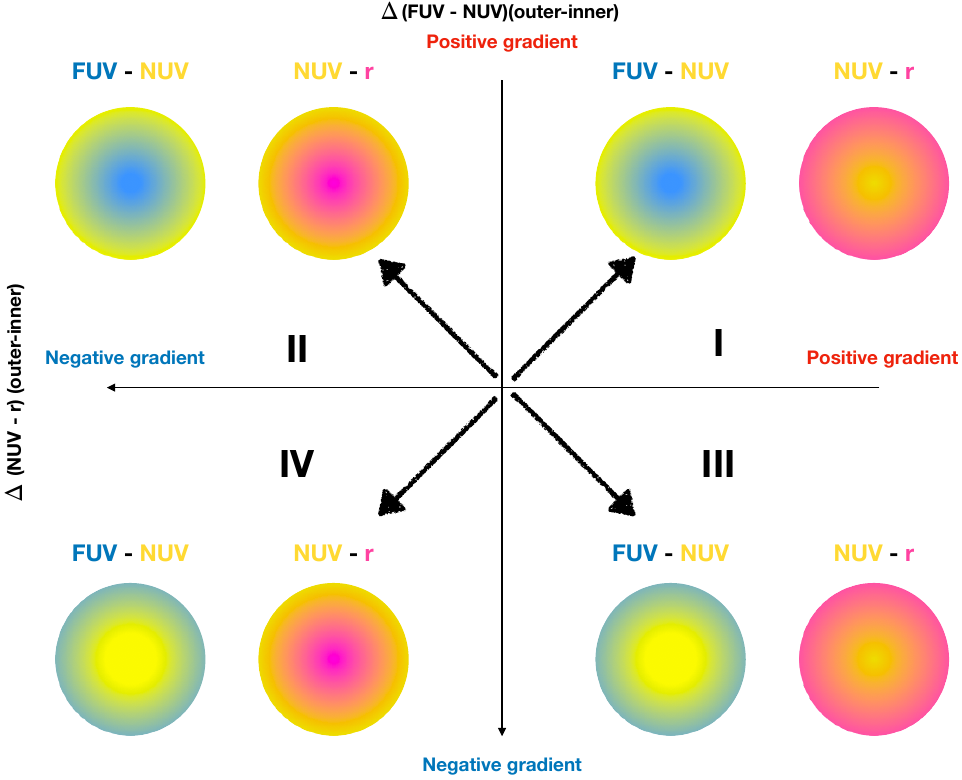}
        \caption{\textbf{Top:} When comparing the color gradient in  $\Delta$FUV-NUV (along the y-axis) and $\Delta$NUV-r (along the x-axis), there are four possible combinations, depending on whether each gradient is positive or negative. We indicate these four quadrants here, labeled I-IV, along with an illustration of the relative color gradients. According to an analysis of very similar colors at $z \sim 0$ from \citet{Abramson+11}, actively or recently ram pressure stripped galaxies should fall in quadrant I. From surveys of field galaxies at $z \sim 2$, showing galaxies having bluer outskirts than insides, the majority of galaxies would lie in quadrant IV, absent significant evolutionary pressure from the environment.}
	\label{fig:quadrant_diagram}
  \end{minipage}%
  \begin{minipage}[b]{1.0\linewidth}
    \centering
    \plotone{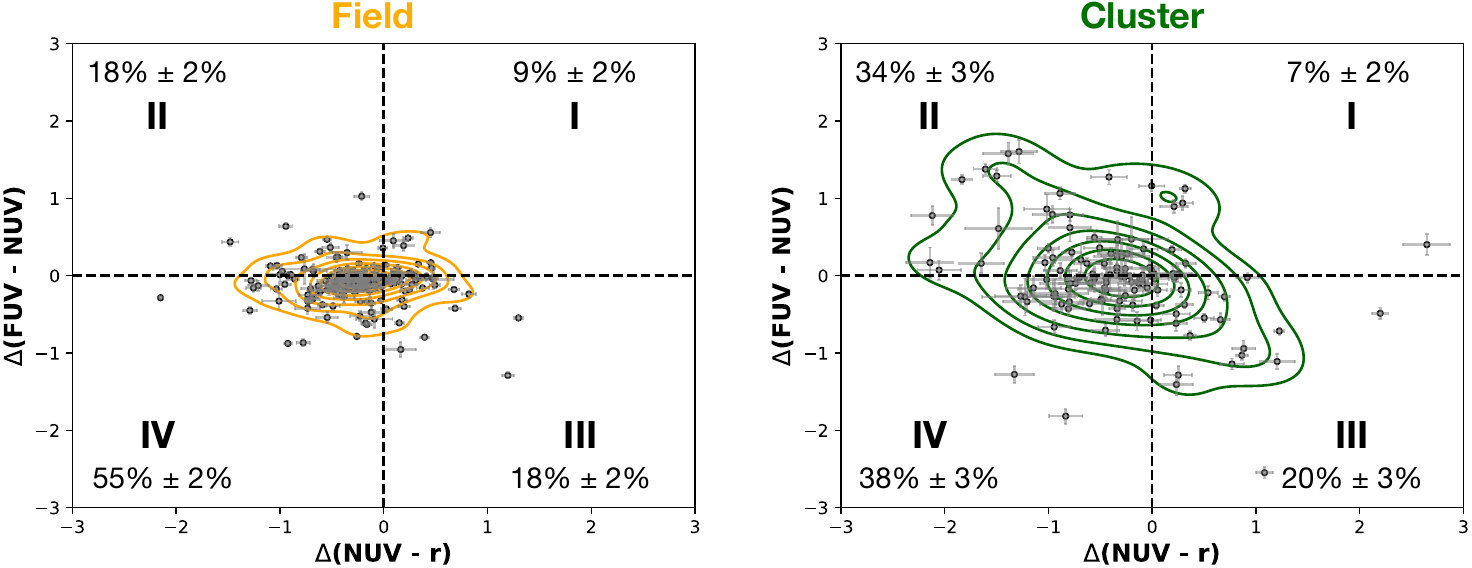}
        \caption{A comparison of the $\Delta$FUV-NUV and $\Delta$NUV-r color gradients of field galaxies (left) and cluster galaxies (right). Axes are in units of AB magnitudes. We also show density contours based on a kernel density estimate (KDE) of the plotted points for another visual representation of the distributions. Overall, both cluster and field have similar amounts of galaxies falling in quadrants I and III. The major difference between cluster and field appears to be the number of galaxies in quadrant IV and quadrant II in the cluster. There is also an overall much wider range in color gradient values in the cluster.}
	\label{fig:quadrant_data}
  \end{minipage}
\end{figure*}

Along with investigating the individual color gradients, studying the combined information for each galaxy of both (FUV$_{\mathrm{outer}}$-NUV$_{\mathrm{outer}}$) - (FUV$_{\mathrm{inner}}$-NUV$_{\mathrm{inner}}$) and (NUV$_{\mathrm{outer}}$-r$_{\mathrm{outer}}$) - (NUV$_{\mathrm{inner}}$-r$_{\mathrm{inner}}$), hereafter referred to as $\Delta$FUV-NUV and $\Delta$NUV-r, has the potential to provide additional insight into galaxy properties. Based on the four possible combinations of gradients for $\Delta$FUV-NUV and $\Delta$NUV-r, which are positive-positive (quadrant I), positive-negative (quadrant II), negative-positive (quadrant III), and negative-negative (quadrant IV) gradients, each falls in a different region of the color gradient quadrant plot, illustrated in Figure \ref{fig:quadrant_diagram} \& \ref{fig:quadrant_data}. Comparing the overall distributions of cluster and field galaxies across these quadrants highlights the differences in galaxy colors driven by the environment.

We find that overall the distribution of cluster galaxies has a larger UV color range, and is more dispersed than field galaxies. The total fraction of galaxies in quadrants I \& III when comparing the cluster and field are similar. The most prominent difference between cluster and field is a $\sim$20\% difference in the fraction of galaxies falling in quadrant IV in the field, versus quadrant II in the cluster.

A galaxy falling in quadrant I, with positive-positive gradients, could be the result of ram pressure stripping of sufficient strength and on a timescale long enough to result in a color difference in the outer aperture, driven by quenching. At low-redshift, \citet{Abramson+11} found that in a known ram pressure stripped galaxy, $\Delta$FUV-NUV and $\Delta$NUV-$r$ were both positive, meaning the galaxy would have redder outskirts than the inner region for both gradients. However, unlike at low-redshift where it is possible to isolate what appear to be dust-free regions, where ram pressure has stripped away all or most of the dust, for our sample at high-redshift the resolution is too poor to make a solid determination of the dust content in the inner or outer apertures with these filters. Thus, an observed positive-positive gradient could also be the result of stronger dust extinction in the outer parts of the galaxy, as well as from stronger inner vs. outer star formation from something like a central starburst. However, \citet{Miller+22} found that only 15$\pm$6\% of star-forming $1.3 < z < 2.0$ galaxies had color gradients consistent with stronger inner than outer star formation, and another 8$\pm$4\% were consistent with stronger outer dust extinction. A significant overabundance of cluster galaxies in this quadrant when compared with the field could thus be evidence for significant enough ram pressure over time to result in quenching. We do not find any significant difference between cluster and field population within quadrant I, indicating this quenching pathway is currently not significant in these clusters.

Quadrant IV, where both $\Delta$FUV-NUV and $\Delta$NUV-r are negative (redder inner colors), is the most commonly observed combination of color gradients in the field comparison sample. \citet{Miller+22} also predicted that the majority of field galaxies would fall in this quadrant as the result of either stronger dust extinction in the center versus the outskirts (69\% of galaxies), or stronger outer star formation (8\% of galaxies).

Quadrants II (where $\Delta$FUV-NUV is positive and $\Delta$NUV-r is negative) and III (where $\Delta$FUV-NUV is negative and $\Delta$NUV-r is positive) are difficult to interpret given the limitations in terms of the color bands we have access to, and the complications of dust extinction. We thus hesitate to speculate on the physical drivers of galaxies being found in these quadrants, and note instead simply that the environment, through any combination of a number of effects like (but not limited to) stripping, strangulation, pre-processing, and tidal interactions, clearly seems to drive an observable difference in the number of galaxies that fall here.

We discuss the implications of the quadrant analysis further in Section \ref{sec:quadrant_disc}.

\section{Discussion}

We find statistically significant differences between the cluster and field population based on the resolved UV/optical color properties. This in and of itself is a significant advancement in our understanding of the influence of the environment at high redshift based on resolved observations.

The interpretation of the physical drivers behind these differences is complicated. This is driven primarily by the difficulty in distinguishing between reddening driven by dust, and/or stellar population age, both of which affect the FUV-NUV and the NUV-r colors in similar ways. There are statistically significant differences in both the inner and outer FUV-NUV colors for cluster and field galaxies, which could be interpreted as either a more quenched population, or increased dust extinction in cluster galaxies. However, there is no overall difference in NUV-r population level colors. FUV-NUV is more sensitive to color changes from recent star formation and dust extinction, indicating that perhaps this wavelength range is necessary to detect differences on the population level between cluster and field at this redshift.

The color gradient distributions are another complicating piece of the overall picture to consider. The overall population level FUV-NUV color gradients of cluster galaxies are significantly different between the cluster and field, with a pronounced redder outside-in tail to the distribution seen in Figure \ref{fig:app_survival_f2}. However, while we do find a difference in the NUV-r color distribution, it does not clearly favoring a particular direction for positive vs. negative color gradients. Clearly, more filters, and rigorous stellar population modelling based on more filters, is needed to further identify the forces within the environment driving these color differences at this redshift.

\subsection{Quadrant Analysis Interpretation}
\label{sec:quadrant_disc}

The quadrant plot clearly shows a difference in the distribution and overall range of color gradients observed between the cluster and field populations. Ram pressure stripped and quenched cluster galaxies are likely to be observed in quadrant I based on low-redshift galaxy studies \citep{Abramson+11, Merluzzi+16, Cramer+19}. Although no large scale studies of color gradients at a population level have been conducted at low or high redshift in clusters, studies of other signatures of ram pressure stripping and quenching have found it affects a very significant number of star forming galaxies in low redshift clusters like Virgo \citep[e.g.][]{Chung+09, Zabel+22, Villanueva+22, Jimenez+23, Zinger+24}. Furthermore, previous studies of galaxies in these three $z \sim 1.6$ clusters found evidence supporting the effects of ram pressure acting on the molecular gas content \citep{Noble+19} and the molecular gas kinematics \citep{Cramer+23}. We expected, given this supporting evidence, that we might find an overabundance of galaxies in quadrant I in the cluster sample. Our analysis, however, finds no significant overabundance of galaxies in this quadrant when compared with the field.

We do, however, find that an additional $\sim$20\% of galaxies in the cluster are found in quadrant II (bluer outside-in NUV-r color gradients, and redder outside-in FUV-NUV color gradients) compared to the field. This could be related to ram pressure interactions observed at low redshift that cause star formation to be triggered in incident regions \citep{Moretti+20, Cramer+20, Cramer+21}. As mentioned previously, the FUV filter tends to trace star formation within stellar populations of ages $\sim10-40$ Myr, while NUV is more sensitive to timescales closer to $\sim$100 Myr \citep{Abramson+11, Ephremova+11, Boqiuen+14}. If ram pressure had triggered star formation, and subsequently cleared out the region of star forming gas resulting in quenching within $\sim 10-40$ Myr as has been observed in some cases at low redshift \citep{Abramson+11, Cramer+20, Cramer+21}, there could be an excess of NUV flux, but a relative deficit of FUV flux.

The presence of some observable effect of the influence of ram pressure would be consistent with results from simulations that show that around $z \sim 2$ \citep{Boselli+22}, ram pressure is only beginning to actively morph galaxies in these clusters. In the case of these clusters, perhaps ram pressure is not yet resulting in significant outside-in quenching that would be seen in quadrant I, but is still influencing the stellar populations in ways like star formation triggering. This could be related to early stages of the development of the ICM near this redshift, that would need to be confirmed with deep X-ray observations of these clusters.

This interpretation is highly speculative. However, the fact remains that, regardless of the physical or observational driver, there are clear differences in the colors and color gradients of cluster galaxies when compared with the field.  Similar large sample-size studies of cluster galaxy color gradients at low redshift, especially in clusters where ram pressure stripped galaxies are easily identified based on previous studies of gas morphology (e.g. the Virgo cluster \citealt{Chung+09, Brown+21}), could be extremely helpful in determining where different physical drivers move galaxies within the quadrant plot. This paper has focused in particular on whether we observe an effect of ram pressure on color gradients in these clusters, but this is not to say that others drivers of evolution in clusters, like strangulation, overconsumption, and gravitational interactions are less important. It is simply a factor of there being a number of published studies of color gradients ranging from the FUV to optical of ram pressure stripped galaxies in the literature with which to compare. Similar color gradient analysis, be it observational or in simulations, of galaxies positively identified as actively experiencing the other above effects would help us to better understand their roles in driving the distributions of galaxies in clusters within the quadrant diagram.

Furthermore, it would also be useful to study whether high-redshift clusters have a different distribution of galaxies within the quadrant plot when compared with low-redshift clusters. This could have direct implications for better understanding the evolution of galaxies within clusters over cosmic time. We hope to implement this sort of color gradient analysis across a range of redshifts in future work, and that this study provides additional motivation for others to pursue these kinds of analyses, to more fully probe the parameter space. 

\subsection{Future Work with JWST}

While the color gradient quadrant analysis of $\Delta$FUV-NUV and $\Delta$NUV-$r$ can be informative for determining physical drivers behind galaxy locations in the diagram, they are also affected by the degenerate effects of dust extinction and quenching. A solution to determining the relative degree to which each is contributing to the observed color differences will need to come from observations that can distinguish between the two scenarios. The \textit{UVJ} diagram is one method for disentangling the effects of dust extinction from differences in stellar population. Galaxies with high dust extinction are reddened in both the U - V and V - J colors \citep{Forrest+16, Martis+16}. Galaxies with older stellar populations are, when compared with the U - V, bluer in the V - J color \citep{Carnall+19, Belli+19}. Moreover, \textit{UVJ} colors have also been shown to correlate with specific star formation rate, allowing further constraints on quenching in cluster vs. field \citep{Leja+19, Akins+22}. At $z >1.6$, the rest-frame $J$ band is redshifted to $>$3 $\mu$m, and therefore JWST NIRCam imaging is the only high-resolution instrument that can provide resolved observations in this regime.

Recent work by \citet{Miller+22}, using a subset of star-forming galaxies ranging from $1.3 < z < 2.0$ from the JWST CEERS survey \citep{Finkelstein+22}, used a combination of JWST filters nearest to the redshifted \textit{UVJ} color bands, and flexible modelling to interpolate to the rest-frame \textit{UVJ} colors, to measure the resolved color profiles for $45$ star-forming field galaxies. By using a similar observing setup with JWST NIRCam as \citet{Miller+22}, a similar color analysis of these cluster galaxies could be easily compared with the field sample. 

\section{Summary}

We have conducted a rigorous statistical analysis of the colors and color gradients, derived from HST observations, of high-redshift galaxies in clusters and in the field. By carefully controlling for confounding variables in these two samples, like filter differences and redshift, differences in these colors and color gradients are informing us about differences in the evolutionary pathway of galaxies in low and high density environments.

\begin{description}

	\item[$\bullet$ Environment drives color differences] We find that cluster galaxies at $z \sim 1.6$ have statistically significantly redder UV colors in inner ($<r_{50}$) and outer ($r_{50} < r < r_{90}$) regions when compared to field galaxies at similar redshift and stellar mass. In contrast, there is no statistically significant difference detected in the NUV-r color in the inner or outer regions of galaxies. Furthermore, cluster galaxies have an overall redder distribution of outside-in FUV-NUV color gradients. This large scale study of the color gradients of cluster galaxies and field galaxies at this redshift is the first of its kind, and reveals clear differences between the two populations.

    \item[$\bullet$ Constraints on outside-in quenching] Previous study of these clusters found evidence consistent with ram pressure stripping of gas \citep{Noble+17, Noble+19, Cramer+23}. However, we find no overabundance of cluster galaxies with outside-in color gradients reflective of quenching driven by prolonged strong ram pressure stripping (like is seen in low redshift galaxies) at this redshift. We do, however, find that an additional $\sim$20\% of galaxies in the cluster are found in quadrant II (bluer outside-in NUV-r color gradients, and redder outside-in FUV-NUV color gradients) compared to the field. This could be consistent with ram pressure influencing the stellar population of galaxies in these clusters through star formation triggering, but merits further investigation.
\end{description}

\acknowledgments

This work was supported from \textit{HST} program number GO-16300, and is based on data and catalog products from HFF-DeepSpace, funded by the National Science Foundation and Space Telescope Science Institute (operated by the Association of Universities for Research in Astronomy, Inc., under NASA contract NAS5-26555).  A.N. additionally acknowledges support from the Beus Center for Cosmic Foundations at Arizona State University, from the National Science Foundation through grant AST-2307877 and through award SOSPA7-025 from the NRAO. GHR acknowledges support from NASA through HST-GO-16300.004, NASA Keck RSA-1699004, from the NSF through AST-2206473, and through NASA via ADAP grant 80NSSC19K0592.  He also acknowledges the generous support from the International Space Sciences Institute (ISSI) via the ``Cosweb" team.  YMB gratefully acknowledges funding from the Netherlands Organization for Scientific Research (NWO) under Veni grant number 639.041.751 and financial support from the Swiss National Science Foundation (SNSF) under funding reference 200021\_213076. JN acknowledges support from the Universidad Andres Bello internal grant DI-07-22/R and Fondecyt Regular \#1230591 (PI Lucia Guaita). GW gratefully acknowledges support from the National Science Foundation through grant AST-2205189 and from HST program number GO-16300. R.D. gratefully acknowledges support by the ANID BASAL project FB210003. Based on observations made with the NASA/ESA Hubble Space Telescope, and obtained from the Hubble Legacy Archive, which is a collaboration between the Space Telescope Science Institute (STScI/NASA), the Space Telescope European Coordinating Facility (ST-ECF/ESA) and the Canadian Astronomy Data Centre (CADC/NRC/CSA).

This research made use of \texttt{photutils}, an Astropy package for detection and photometry of astronomical sources \citep{photutils}. Many thanAD to the lead developer of \texttt{photutils}, Larry Bradley, for his help in developing additional code for our particular need.

Thank you to Zhongxue Chen for his and his group's help in  advising us on survival statistics techniques. Also thank you to the anonymous referee for their very helpful comments and suggestions. 

\software{\texttt{SwARP} (version: 2.19.1; \citet{Bertin+02}), Source Extractor (version 2.24.2; \citep{Bertin+96}, \texttt{EzGal} \citep{Mancone+12}, \texttt{stsynphot} \citep{stsynphot}, \texttt{photutils} \citep{photutils}, \texttt{FAST} \citep{FAST}}

\clearpage

\bibliography{Bibliography}{}
\bibliographystyle{aasjournal}

\clearpage

\appendix

\restartappendixnumbering

\section{Full photo-z and spec-z cluster galaxy sample vs. just spec-z}

\begin{figure*}
\plotone{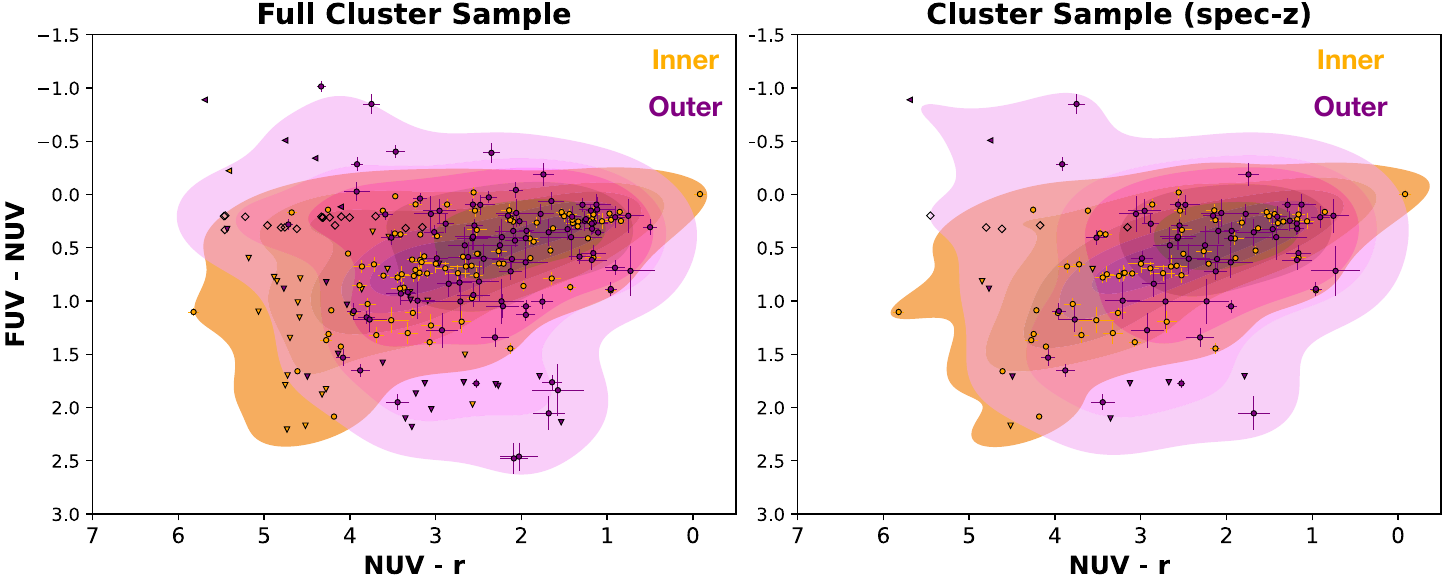}
	\caption{A comparison of the inner (orange) and outer (pink) colors in NUV-r vs. FUV-NUV for the full (photo and spec-z) cluster sample used in the analysis from the main text (left), as well as a spec-z only sample (right). Axes are in units of AB magnitudes. Right facing triangles show aperture fluxes with limits in either the FUV flux, or the NUV flux. Downward facing arrows show limits in just the NUV flux. Open circles show aperture fluxes with limit fluxes in both FUV and NUV. Contours based on a kernel density estimate (KDE) are also shown to better illustrate the comparison.}
	\label{fig:app_specz_colors}
\end{figure*}

\begin{figure*}
\plotone{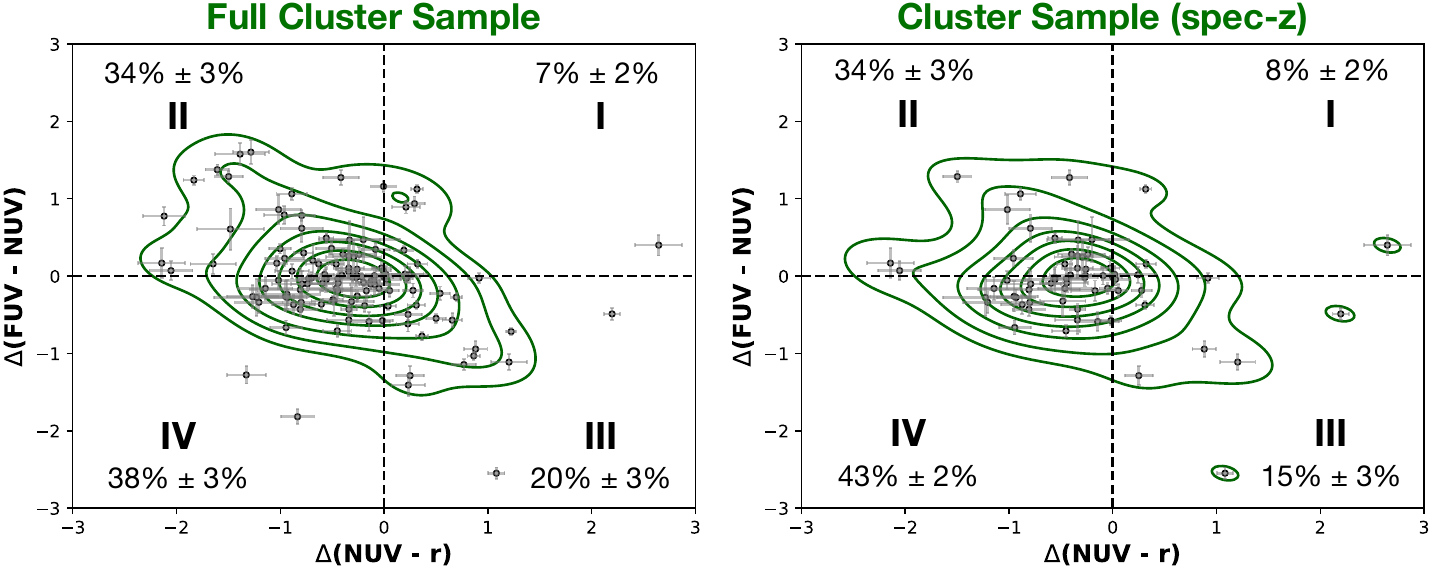}
	\caption{A comparison of the $\Delta$FUV-NUV and $\Delta$NUV-r color gradients for the full (photo and spec-z) cluster sample used in the analysis from the main text (right), as well as a spec-z only sample (left). Axes are in units of AB magnitudes. We also show contours based on a kernel density estimate (KDE) of the plotted points for another visual representation of the distributions. Overall, both samples have similar distributions across all four quadrants, as well as equal representation in each quadrant within the estimated error from a Monte Carlo Markov Chain.}
	\label{fig:app_specz_quadrant}
\end{figure*}

To assess whether there is any significant risk to our analysis from contamination of non-cluster members based on poorly constrained redshift estimates from photo-$z$ methods, we compare with a rerun analysis using a sample containing only galaxies with a spectroscopic redshift between $1.53 < z < 1.69$. This restricted sample contains $\sim$50\% of the larger sample used in the main text. As shown in Figures \ref{fig:app_specz_colors} \& \ref{fig:app_specz_quadrant}, there is no significant difference in the distribution of aperture fluxes in the color diagram, or in the distribution of color gradients in the quadrant plot. Furthermore, a reproduced survival analysis found no difference in the \textit{p}-value significance of any of the 6 tested parameters. Thus, we are confident that our conclusions are not impacted by contamination from uncertain photometric redshift estimates.

\section{Survival Statistics Error Estimation}

\begin{figure*}
\plotone{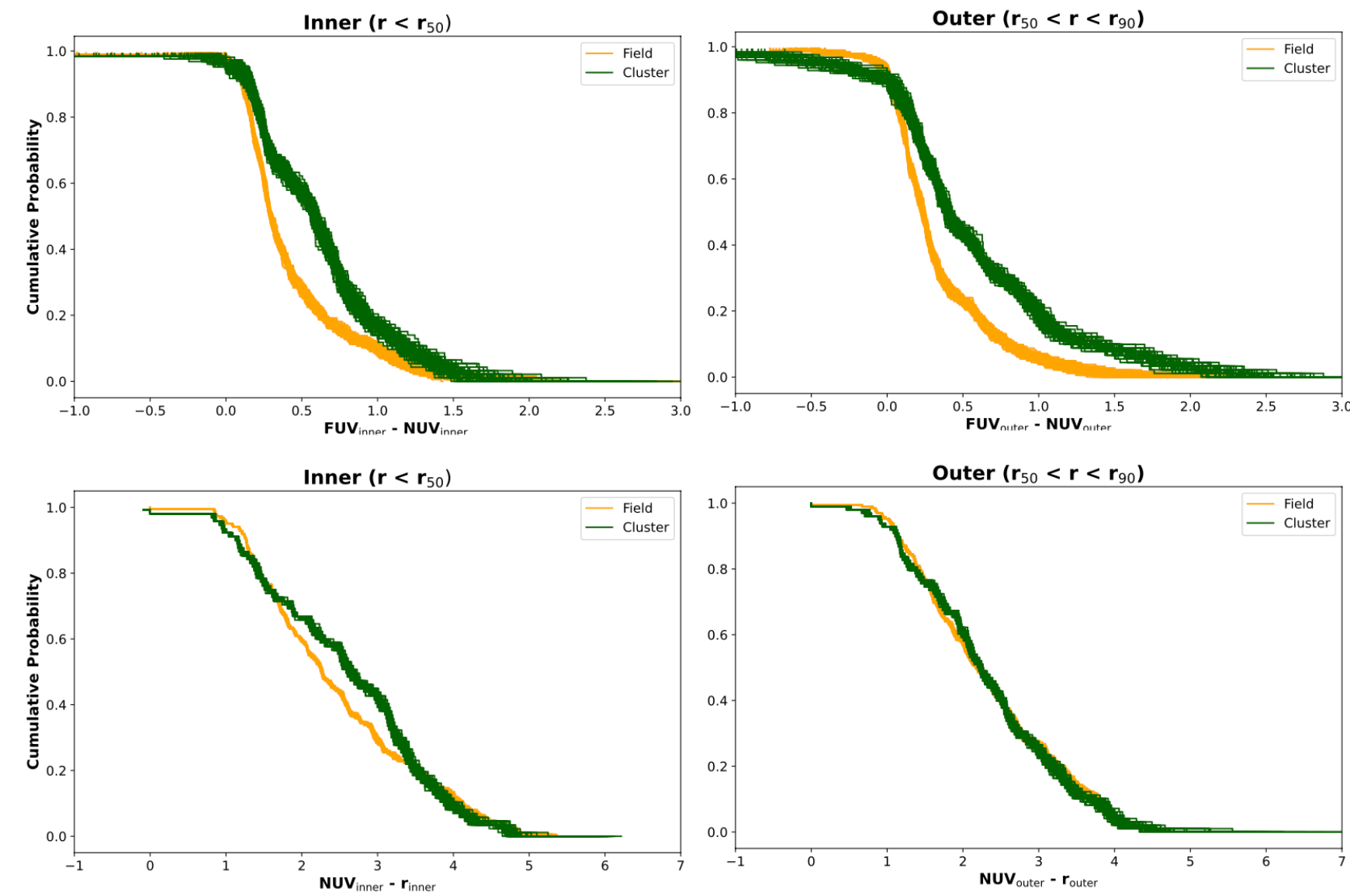}
	\caption{Here we show the cumulative distribution functions of the inner region of cluster galaxies (blue) and field galaxies (orange) in the FUV-NUV, and NUV-r, inner and outer color, generated with survival statistics. Survival statistics does not incorporate errors on the input values directly. Thus, to test the robustness of our results, we run a Monte Carlo, with 100 iterations within the 1$\sigma$ flux errors, of the survival statistics cumulative distribution function for both cluster and field. All iterations are plotted here. We then calculate the rms of the \textit{p}-value for each pair of Monte Carlo generated distributions, in order to calculate an error estimate for the \textit{p}-value.}
	\label{fig:app_survival_f1}
\end{figure*}

\begin{figure*}
\plotone{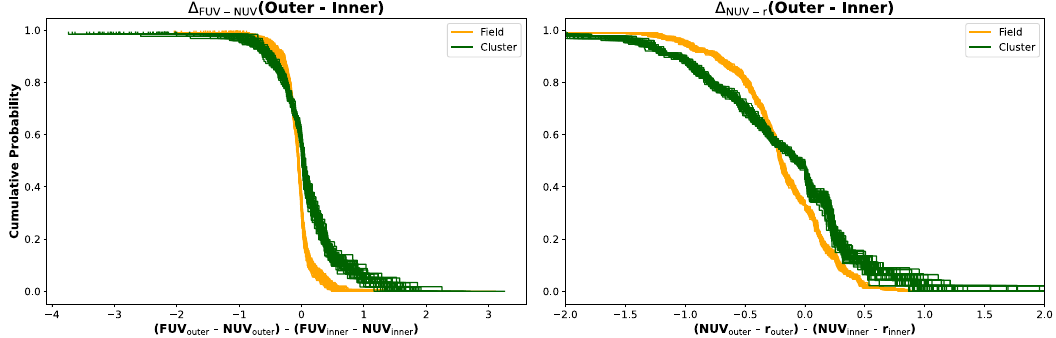}
	\caption{Same caption as Figure \ref{fig:app_survival_f1}, but for the $\Delta$FUV-NUV and $\Delta$NUV-r color gradients.}
	\label{fig:app_survival_f2}
\end{figure*}

Survival analysis does not directly incorporate errors in its Kaplan-Meier fit, or a log-rank comparison of those fits. In order to assess the impact of these errors on the \textit{p}-value we calculate, we implement a Monte-Carlo approach, using the 1$\sigma$ errors from our empty aperture analysis. We vary each filter component of each color within the 1$\sigma$ error, then proceed with a Kaplan Meier fit, and calculate a \textit{p}-value with a log-rank method, over 100 iterations. This allows us to estimate the variance of the \textit{p}-value within the 1$\sigma$ flux errors. These Kaplan Meier fits are shown in Figures \ref{fig:app_survival_f1} \& \ref{fig:app_survival_f2}. We find that the individual flux errors do not contribute to a significant change in the overall confidence level of the \textit{p}-values we calculate. 

\end{document}